\documentclass[prd,twocolumn,aps,nofootinbib,showpacs]{revtex4}
\usepackage{graphicx}
\usepackage{dcolumn}
\usepackage{bm}
\usepackage{amsmath}
\usepackage{amsfonts}
\usepackage{amssymb}
\usepackage{appendix}
\usepackage{pdfpages}
\usepackage{graphicx}
\usepackage{wrapfig}
\usepackage{multirow}
\usepackage{mathrsfs}
\usepackage{epstopdf}
\usepackage{slashed}
\usepackage{soul}
\usepackage{mathtools}
\usepackage{floatrow}
\usepackage{float}
 \usepackage{csquotes}
\usepackage{wrapfig}
\usepackage{graphicx}
\usepackage{epsfig}
\usepackage{hyperref}
\usepackage[utf8]{inputenc}
\usepackage{epsfig}
\usepackage{dcolumn}
\usepackage{morefloats}

\def\be{\begin{equation}}
\def\ee{\end{equation}}
\def\bea{\begin{eqnarray}}
\def\eea{\end{eqnarray}}

\def\gsim{\ \rlap{\raise 2pt\hbox{$>$}}{\lower 2pt \hbox{$\sim$}}\ }
\def\lsim{\ \rlap{\raise 2pt\hbox{$<$}}{\lower 2pt \hbox{$\sim$}}\ }
\def\dslash{\kern-4pt \not{\hbox{\kern-2pt $\partial$}}}
\def\pslash{\not{\hbox{\kern-2pt p}}}

\def\infinity{\rotatebox{90}{8}}
\newcommand{\Eps}{\epsilon}

\newcommand{\dcp}{\delta_{CP}}

\begin{document}
\DeclareGraphicsExtensions{.eps,.ps}
\title{\boldmath Can nonstandard interactions jeopardize the hierarchy sensitivity of DUNE ? }
\author{K. N. Deepthi}
\email[Email Address: ]{deepthi@prl.res.in}
\affiliation{
Physical Research Laboratory, Navrangpura,
Ahmedabad 380 009, India}

\author{Srubabati Goswami}
\email[Email Address: ]{sruba@prl.res.in}
\affiliation{
Physical Research Laboratory, Navrangpura,
Ahmedabad 380 009, India}

\author{Newton Nath}  
\email[Email Address: ]{newton@prl.res.in}
\affiliation{
Physical Research Laboratory, Navrangpura,
Ahmedabad 380 009, India}
\affiliation{Indian Institute of Technology, Gandhinagar, Ahmedabad--382424, India}
\begin{abstract}
We study the effect of non-standard interactions (NSIs) 
on the propagation of neutrinos through the Earth matter and how it affects the hierarchy sensitivity of the DUNE experiment. We emphasize on the special case
when the  diagonal NSI parameter 
$\epsilon_{ee} = -1$, nullifying the standard matter effect. 
We show that, if in addition, CP violation is maximal then this gives rise to
an exact intrinsic hierarchy degeneracy in the appearance channel, 
irrespective of the  baseline and energy. Introduction of  off-diagonal NSI parameter, $\epsilon_{e \tau}$,  shifts the position of this degeneracy to a
different $\epsilon_{ee}$. Moreover the unknown magnitude and  phases of  
the off-diagonal NSI parameters can give rise to additional degeneracies. 
Overall, given the current model independent limits on NSI parameters,  
the hierarchy sensitivity of DUNE can get seriously impacted. 
However, a more precise knowledge on the NSI parameters, specially $\epsilon_{ee}$, 
can give rise to an improved sensitivity. 
Alternatively, if NSI exists in nature,
and still DUNE  shows hierarchy sensitivity,  
certain ranges of the NSI parameters can be excluded. Additionally, we briefly discussed the implications 
of $\epsilon_{ee} = -1$ (in the Earth) on MSW effect in the Sun.
\end{abstract}
\maketitle
\textbf{Introduction:} Phenomenal experiments over the past 
decades have established neutrino oscillations and led us into an era of 
precision measurements in the leptonic sector. Current data determines
the two mass squared differences ($\Delta m^2_{21} = m_2^2 -m_1^2$, $|\Delta m^2_{31}|=|m_3^2 - m_1^2| $,
$m_1,m_2,m_3$ being the mass states) and three leptonic mixing angles
($\theta_{12}$, $\theta_{23}$, $\theta_{13}$) 
with considerable precision \cite{Capozzi:2016rtj}.
This leaves determination of neutrino mass hierarchy i.e. 
whether  $m_3 > m_2 > m_1$ (normal hierarchy (NH))
or $m_3 < m_1 \approx m_2$ (inverted hierarchy (IH)), 
octant of $\theta_{23}$ i.e  whether $\theta_{23} < \pi/4$ and lies in lower
octant (LO) or it is $>\pi/4$ and is in the higher octant (HO)  
and measurement of $\delta_{CP}$ as the major
objectives of ongoing and future experiments.
Recently, the on-going T2K  experiment 
\cite{Abe:2013hdq}
and global analysis of data 
\cite{Capozzi:2016rtj} 
have hinted that the  Dirac CP phase is maximal i.e. $\dcp = - \pi/2$
although at $3\sigma$ the 
full range ($0-2\pi$) remains allowed. Whereas, recent NO$\nu$A result \cite{Adamson:2017gxd} 
suggested that there are two best fit points if neutrinos obey normal hierarchy, 
(i) $\sin^{2}\theta_{23} = 0.404$, $\delta_{CP}=1.48\pi (\sim -90^\circ)$ and (ii) $\sin^{2}\theta_{23} = 0.623$, $\delta_{CP}=0.74\pi (\sim 135^\circ)$. Also, inverted mass hierarchy with $\theta_{23}<45^\circ$ is disfavoured at 93$\%$ C.L. for all values of $\delta_{CP}$.

    Although neutrino oscillation has been identified as the dominant 
phenomenon to explain the results of various experiments, 
the possibility of sub-leading effects originating from 
\textit{new physics} beyond the Standard Model (SM) cannot be ignored.
Among these, non-standard interactions  have received a lot of 
attention lately specially 
with the emergence of proposed 
next generation experiments like DUNE,T2HK, T2HKK etc. 
\cite{Fukasawa:2016gvm,Coloma:2015kiu,Miranda:2004nb, Agarwalla:2016fkh,Blennow:2016etl,Liao:2016hsa,
Coloma:2016gei,Masud:2016bvp,Masud:2016gcl,Fukasawa:2016lew,deGouvea:2015ndi}.

\indent	\emph{In this work, we emphasize on an interesting case when
standard matter effects during neutrino propagation through the Earth matter  gets nullified 
due to NSI effects \footnote{We neglect the production and detection NSI,   
bounds on which are stronger  by an order of magnitude than matter NSI \cite{Biggio:2009nt}.}, creating degeneracies which affect the 
determination of hierarchy in any long-baseline (LBL) experiment.   
}
Though additional hierarchy degeneracy in LBL experiments due
to NSI have been discussed earlier 
\cite{GonzalezGarcia:2011my,Coloma:2016gei} 
the new  points that we make are : 
(i) if the NSI parameter
$\epsilon_{ee}$  
characterizing new interactions between
electrons  neutrinos and 
electrons has the value  
$\epsilon_{ee} = -1$, the NSI effect cancels the standard matter effect;
(ii) if in addition  
$\delta_{CP} = \pm \pi/2$,
there exists an exact intrinsic hierarchy degeneracy  
in the appearance channel, which is independent of baseline and the 
neutrino beam energy making it unsolvable in LBL experiments, in particular 
DUNE. 
Note that this degeneracy cannot be lifted even if 
$\epsilon_{ee}$ and $\delta_{CP}$ are precisely measured around these values.
This result assumes more importance in the light of current data from the 
T2K experiment
hinting at  $\delta_{CP} \sim -\pi/2$. 
We also consider the simultaneous presence of 
$\epsilon_{ee}$ and non-zero values of the  NSI parameter 
$\epsilon_{e\tau}$ and show that the intrinsic degeneracy still 
exists, albeit at a different $\epsilon_{ee}$. 
Moreover, given the current model independent bounds on the NSI parameters,
hierarchy sensitivity in DUNE does not improve. Rather the phases 
associated with $\epsilon_{e \tau}$ can give rise to additional 
degeneracies which seriously impact the sensitivity. 

\indent We focus on the matter NSI effects on the 
neutrino propagation. This is described by dimension-six four-fermion operators of the form~\cite{Wolfenstein:1977ue} 

\setlength{\abovedisplayskip}{5pt}
\setlength{\belowdisplayskip}{5pt}  
\be
  \label{eq:NSI}
  \mathcal{L}^{NC}_\text{NSI} =(\overline{\nu}_\alpha \gamma^{\rho} P_L \nu_\beta )
        ( \bar{f} \gamma_{\rho} P_C f )2\sqrt{2}G_F
   \epsilon^{fC}_{\alpha\beta} + \text{h.c.}
\ee
where $\epsilon^{f C}_{\alpha\beta}$ are NSI parameters, $\alpha, \beta = e, \mu, \tau$, $C = L,R$ denotes the chirality, $f = u,d,e$, and $ G_{F} $ is the Fermi constant. The Hamiltonian in the flavor basis  can be written as,
\be \label{nsi_hamil}
H = \frac{1}{2E} \left[ U\text{diag}(0,\Delta m^2_{21},\Delta m^2_{31})
U^\dagger + V\right]\,,
\ee
where $U$ is the PMNS mixing matrix having three mixing angles ($ \theta_{ij},~i<j=1,2,3 $) and a CP phase $ \delta_{CP} $. 
$V$ is  the matter potential due to the  the interactions of neutrinos,
\be
V = A \left(\begin{array}{ccc}
1 + \epsilon_{ee} & \epsilon_{e\mu}e^{i\phi_{e\mu}} & \epsilon_{e\tau}e^{i\phi_{e\tau}}
\\
\epsilon_{e\mu}e^{-i\phi_{e\mu}} & \epsilon_{\mu\mu} & \epsilon_{\mu\tau}e^{i\phi_{\mu\tau}}
\\
\epsilon_{e\tau}e^{-i\phi_{e\tau}}& \epsilon_{\mu\tau}e^{-i\phi_{\mu\tau}} & \epsilon_{\tau\tau}
\end{array}\right)\,,
\label{eq:potential}
\ee
where,
$A \equiv 2\sqrt2 G_F N_e(r) E$. The unit contribution to the (1,1) element of the $V$ matrix is the usual matter term arising due to the standard charged-current interactions. Here, diagonal elements of the $V$ matrix are real due to the Hermiticity of  the Hamiltonian in eq.~(\ref{nsi_hamil}). 
$\epsilon_{\alpha\beta}e^{i\phi_{\alpha\beta}}\equiv\sum\limits_{f,C}\epsilon^{f C}_{\alpha\beta}\frac{N_f(r)}{N_e(r)}$, with $N_f(r)$ being the number density of fermion $f$ along the trajectory `$r$' in the matter.
$\epsilon_{\alpha\beta}$ can be written in terms of $Y_n(r)= \frac{N_n(r)}{N_e(r)}$ for neutral matter as
\begin{equation}
\epsilon_{\alpha\beta}=\epsilon_{\alpha\beta}^e+ (2+Y_n(r)) \epsilon_{\alpha\beta}^u+ (1+2 Y_n(r))\epsilon_{\alpha\beta}^d
\label{eq:eps-ab}
\end{equation}
Eq.~(\ref{eq:eps-ab}) is a general equation valid for both Earth and the Sun.
Usually, the symbols $\epsilon_{ee}^\oplus$, $\epsilon_{ee}^\odot$ are 
used to represent  $\epsilon_{\alpha\beta}$ for the matter composition 
of the Earth and the Sun respectively. 
However, for  convenience we have used the 
notations $\epsilon_{ee}^\oplus = \epsilon_{ee}$ in the Earth and $\epsilon_{ee}^\odot$ in the Sun, throughout the paper.\\
\indent For the case with only diagonal NSI parameter, 
we observe from eq.~(\ref{eq:potential}) that for 
$\epsilon_{ee}=-1$, the NSI effect cancels the standard matter effect.
The relevant appearance channel  probability ($P_{\mu e}$)
for normal hierarchy (NH), can be expressed 
in terms of  diagonal NSI parameter $\epsilon_{ee}$ and 
small parameters, $s_{13}$, $r = \Delta m^2_{21}/\Delta m^2_{31}$ and the off-diagonal NSI parameter  $\epsilon_{e \tau}$ as, \cite{Liao:2016hsa},

\begin{widetext} {\small
\bea
P_{\mu e} &=& x^2 f^2 + 2xyfg \cos(\Delta + \delta_{CP}) + y^2 g^2 + \mathcal{O}(\epsilon_{e\mu})
\nonumber\\
&+& 4\hat A \epsilon_{e\tau} s_{23} c_{23}
\left\{ xf [f \cos(\phi_{e\tau}+\delta)  
- g \cos(\Delta+\delta+\phi_{e\tau})] \right.
 \left. -yg [g \cos\phi_{e\tau} - f \cos(\Delta-\phi_{e\tau})]\right\}
\nonumber\\
&+& 4 \hat A^2 g^2 c_{23}^2 | s_{23}\epsilon_{e\tau}|^2
 +  4 \hat A^2 f^2 s_{23}^2 |c_{23}\epsilon_{e\tau}|^2
- 8 \hat A^2 fg s^{2}_{23} c^{2}_{23} \epsilon_{e\tau}^2 \cos\Delta 
+ {\cal O}(s_{13}^2 \epsilon_{e \tau}, s_{13}\epsilon_{e \tau}^2, \epsilon_{e \tau}^3)
\label{eq:prob}
\eea
\bea
&& x=2 s_{13} s_{23},~ y=r  c_{23} \sin 2\theta_{12},~
\Delta=\frac{\Delta m^2_{31} L}{4E},~ \hat A=\frac{A}{\Delta m^2_{31}},
f,\bar{f} = \frac{\sin[\Delta(1\mp\hat A(1+\epsilon_{ee}))]}{(1\mp\hat A(1+\epsilon_{ee}))}\,,\ 
g = \frac{\sin[\hat A(1+\epsilon_{ee}) \Delta]}{\hat A(1+\epsilon_{ee})}
\label{eq:define}
\eea }
\end{widetext}

In the above expressions $s_{ij} = \sin\theta_{ij},c_{ij}=\cos\theta_{ij}, i<j,i,j=1,2,3$.
The expressions for IH can be obtained by replacing 
$\Delta m^2_{31} \to - \Delta m^2_{31}$ (implying $ \Delta \to - \Delta$,  $\hat A \to - \hat A$ 
(i.e. $f \to - \bar{f}$ and $g \to -g$), $ y \to -y $ ). 
Similar expressions for  anti neutrino probability 
($ P_{\overline{\mu} \overline{e}} $) can be obtained by replacing 
 $\hat A \to - \hat A$ (i.e. $f \to \bar{f}$),
$\delta_{CP} \to - \delta_{CP}$, $ \phi_{\alpha \beta} \to - \phi_{\alpha \beta} $. 
From eq.~(\ref{eq:define}) the 
only diagonal parameter to which appearance channel is sensitive is  
$ \epsilon_{ee} $.
Also note that since $P_{\mu e}$ contains 
terms of the form $\sin 2\theta_{12}$ the $s_{12} \leftrightarrow c_{12}$ transformation is not very important in our discussion and we concentrate 
on the region $\theta_{12} <  \pi/4$ \cite{Miranda:2004nb}.    

\indent While assuming the presence of only the diagonal NSI parameters if $\epsilon_{ee}=-1$
, we observe from eq.~(\ref{eq:potential}) that the NSI effect cancels the standard matter effect and
$P_{\mu e}$, for NH,   
can be expressed as, 
\footnote{At $\epsilon_{ee} = -1$, $ f = \sin\Delta (- \sin\Delta) $ and $ g = 1 (-1) $ for NH (IH).}  
\begin{equation} 
P_{\mu e} = x^2 \sin^2\Delta +y^2 + 2 x y g \sin \Delta  \cos(\dcp +\Delta)\,.
\end{equation}
This has a $\Delta \rightarrow -\Delta$ and $\dcp \rightarrow \pi - \dcp$ 
degeneracy. If $\dcp$ is measured accurately such that  $\pi  - \dcp$ is 
not allowed then this degeneracy can be alleviated. However for    
true value of  $\dcp = \pm \pi/2$ since both $\dcp$ and $\pi - \dcp$ are same 
there is an intrinsic degeneracy  which cannot be 
removed even if $\dcp$ is measured precisely around these values. 
Note that for $\dcp = \pm \pi/2$, 
the third term
becomes $ \sin^{2}\Delta $ and there is no hierarchy sensitivity.
Note that this is a special case of the $\epsilon_{ee} \rightarrow 
-\epsilon_{ee} - 2$ and $\delta_{CP} \rightarrow \pi-\delta_{CP}$ 
 degeneracy discussed in  
\cite{Coloma:2016gei}. 
We emphasize on the intrinsic nature of this degeneracy for
$\epsilon_{ee} = -1$ \footnote{In \cite{Yasuda:2015lwa}, a fit to the SuperKamiokande data assuming NSI gives the best-fit as $\epsilon_{ee}^\oplus = -1$, $|\epsilon_{e \tau}^\oplus|=0$. However, the author mentions that this
could be because of the discrepancies between the Monte Carlo simulation of the Superkamiokande group and theirs.} and $\dcp = \pm \pi/2$. 
\newpage
\par
\begin{figure}[H]
\begin{center}
 \begin{tabular}{lr}
 \includegraphics[height=5cm,width=5.7cm]{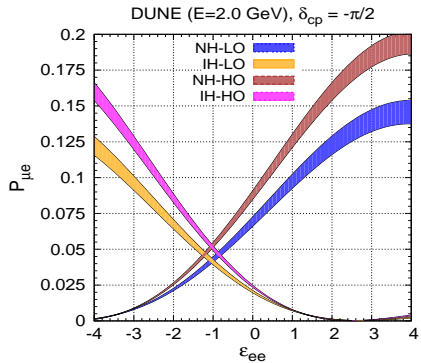} \\ 
 \end{tabular}
 \end{center}
\vspace{-4ex}        
\caption{$P_{\mu e}  $  vs $ \epsilon_{ee} $ for DUNE. 
The bands are over  $\theta_{23}$, for LO($\theta_{23} = 39^\circ - 42^\circ$) and for HO($\theta_{23} = 48^\circ - 51^\circ$).}
\label{dune_NSI_deg}
\end{figure}

 The behaviour of $P_{\mu e}$  as a function of $ \epsilon_{ee} $ is shown in fig.~(\ref{dune_NSI_deg}) for a fixed energy E = 2 GeV and 
$\dcp = - \pi/2$ for DUNE experiment.
The 4 bands correspond to different combinations of 
hierarchy and octant as labelled in the figure.
The width of the bands are due to 
variation over $\theta_{23}$. 
We observe  that the  probability is a rising (falling)
function of $ \epsilon_{ee} $ for NH (IH) for both the octants.
 The value $ \epsilon_{ee} = 0$, represents the standard oscillation case
for which a huge separation between NH and IH bands can
be seen for neutrinos,
as the  DUNE baseline has large matter effect.  

But in presence of NSI, DUNE loses hierarchy sensitivity due to additional degeneracy creeping in through the NSI parameters. 
The figure exhibits the   
$\epsilon_{ee} \rightarrow -\epsilon_{ee} - 2$  degeneracy
discussed in \cite{Coloma:2016gei}. 
One further observes that, for $\epsilon_{ee} = -1$, the 
bands for NH-LO(NH-HO) overlap with those of IH-LO(IH-HO) demonstrating the  
intrinsic nature of the hierarchy degeneracy.

It's also noteworthy that this degeneracy will be there for 
all baselines and hence no LBL experiment will be able to 
resolve this degeneracy. 
We have checked that similar plots for 
antineutrinos show that  the nature of the octant degeneracy as a function of $ \epsilon_{ee} $ is opposite 
and hence, running in ($\nu+\overline{\nu}$)
mode will take care of the wrong octant solutions. 
For known octant, the  probability figures show the absence of 
hierarchy degeneracy for $ \epsilon_{ee} > 2 $.

\begin{figure}[H] 
\begin{center}
   \includegraphics[height=5cm,width=5.7cm]{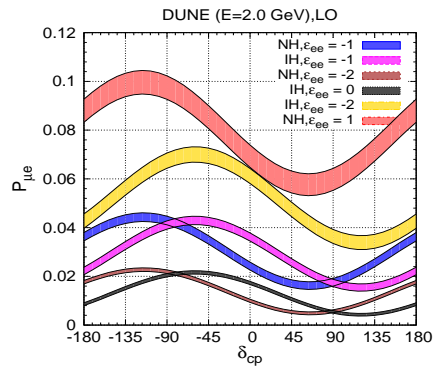}
\end{center} 
\vspace{-4ex}
\caption{$ P_{\mu e} $  vs $ \dcp $ for DUNE. The bands are for $ \theta_{23} \in $  LO.}
 \label{fig:dune_lo}
\end{figure}

So far we  have discussed hierarchy degeneracy for $\dcp =
- \pi/2$. 
 In fig.~(\ref{fig:dune_lo}), we have plotted the  probability vs $ \dcp $ for
various values
of $ \epsilon_{ee} $ to understand the degeneracies due to $\dcp$
in presence of NSI. 
The various degenerate solutions observed are,
(i) The WH-RO-R$\dcp$\footnote{Note that  WH = Wrong Hierarchy, RO = Right Octant, R$\dcp$ = Right $\dcp$, W$\dcp$ = Wrong $\dcp$.} solution, discussed above, 
occurs at $\dcp = \pm \pi/2$.
This is seen by comparing the blue and magenta bands 
or the brown and the grey bands.  
Mathematically the above implies   $P^{NH}_{\mu e}(\epsilon_{ee}, \dcp = \pm \pi/2 ) = P^{IH}_{\mu e}(- \epsilon_{ee} - 2, \dcp=\pm \pi/2)$.
(ii) The  WH-RO-W$\dcp$  solutions which can be observed by comparing
the blue band with magenta band or brown band with grey band by drawing a horizontal line 
corresponding to a given probability.
This can be defined as $P^{NH}_{\mu e}(\epsilon_{ee}, \dcp) = P^{IH}_{\mu e}(-\epsilon_{ee} - 2, \dcp^{\prime})$.
(iii) Apart from the degenerate solutions corresponding to $\epsilon_{ee} \rightarrow -\epsilon_{ee} -2$  one can have a more general form of the degeneracy $P^{NH}_{\mu e}(\epsilon_{ee}, \dcp) = P^{IH}_{\mu e}(\epsilon^{\prime}_{ee}, \dcp^{\prime})$.
This can be seen by comparing the yellow band with the dark-red band. 
The conclusions made above are also true for antineutrinos.
%

\textbf{Off-diagonal NSI :}
We also study the cases of  whether inclusion of off-diagonal NSI parameter 
$ \epsilon_{e \tau} $ can  resolve  the intrinsic hierarchy degeneracy 
occurring at $ \epsilon_{e e} = -1 $  and $\dcp = -\pi/2$.  
\footnote{Since, the bounds on $\epsilon_{e \mu}$ are more tightly constrained than $\epsilon_{e \tau}$ we consider the effect of latter.}
From eq.~(\ref{eq:prob}) 
the difference in hierarchy for $ \epsilon_{e e} = -1 $
and $\dcp = -\pi/2$ and same $\epsilon_{e \tau}$
in both NH and IH can be expressed as, \\
\begin{align} \label{eq:eps_etau_ex}
P^{NH}_{\mu e} - P^{IH}_{\mu e} & = 8 \hat A \sin \Delta c_{23} s_{23}  (x \sin \Delta  -  x \cos \Delta \nonumber \\
& +  y  \sin \Delta ) \epsilon_{e \tau} \sin \phi_{e \tau} 
\end{align}

\noindent Thus, we can see that  
there is a finite 
difference between the NH and IH probabilities which vanishes if 
$ \phi_{e \tau} $  is zero or if $\phi_{e\tau} \rightarrow -\phi_{e\tau}$. 
However, the intrinsic degeneracy 
may shift   
by an amount which depends on the values of 
the off-diagonal NSI parameters. For instance, assuming
a small shift `q'
in presence of $\epsilon_{e \tau}$ one can write  
$P^{NH}_{\mu e}(\epsilon_{ee} + q, \epsilon_{e \tau}, \dcp = \pm \pi/2) = P^{IH}_{\mu e}(\epsilon_{ee} + q,  \epsilon_{e \tau},\dcp=\pm \pi/2)$. Then
assuming $\delta = -\pi/2$ and $\theta_{23} = \pi/4$, $ q $
can be calculated at the oscillation maxima ($\sin\Delta \sim \pi/2$) as, 

\begin{eqnarray} \label{eq:etau_shift}
q =  -\frac{0.23 \sin\phi_{e \tau} \epsilon_{e \tau}}{0.046-0.03 \cos \phi_{e \tau} + 0.01 \epsilon^{2}_{e \tau}} 
\end{eqnarray}

\begin{figure}[H] 
\begin{center}
\includegraphics[height=5cm,width=5.7cm]{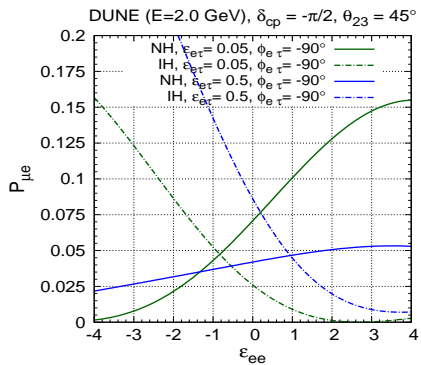}
\end{center} 
\vspace{-4ex}
\caption{ $ P_{\mu e} $  vs  $ \epsilon_{ee} $ for different  $ \epsilon_{e \tau} $ values.}
\label{fig:prob_ee_45}
\end{figure}

For instance, for $\epsilon_{e \tau} = 0.05(0.5)$ and
$\phi_{e \tau} = -90^\circ$, 
eq.~(\ref{eq:etau_shift}) gives $q \simeq 0.25(2.37)$. 
This shift can be observed from the green(blue) dashed and solid lines of fig.~(\ref{fig:prob_ee_45}).
These two cases demonstrate degeneracy of the form
$P^{NH}_{\mu e}(\epsilon_{ee}, \dcp, \epsilon_{e \tau}, \phi_{e \tau}) = P^{IH}_{\mu e}(\epsilon_{ee}, \dcp, \epsilon_{e \tau},\phi_{e \tau})$ with $\epsilon_{ee}\neq -1$ and $\dcp = \pm \pi/2$.

\par Additionally, we can also observe more general degeneracy of the form $P^{NH}_{\mu e}(\epsilon_{ee}, \dcp, \epsilon_{e \tau}, \phi_{e \tau}) = P^{IH}_{\mu e}(\epsilon_{ee}^\prime, \dcp, \epsilon_{e \tau}^\prime, \phi_{e \tau}$) by drawing  horizontal lines
intersecting the green(blue)-dashed and blue(green)-solid lines.  
Fig~(\ref{fig:prob_ee_45}) is drawn for fixed values of $\dcp$ and $\epsilon_{e \tau}$. Allowing these parameters to vary can generate additional degeneracies. 

Note that eq~(\ref{eq:etau_shift}) and fig.~(\ref{fig:prob_ee_45}) 
correspond to the energy at which the oscillation maxima occurs. 
We have verified that in general
the amount of this shift depends on the energy and   
for a different energy the  intrinsic
degeneracy occurs at a different value of $\epsilon_{ee}$. 
Thus the intrinsic degeneracy  becomes energy dependent in presence
of $\epsilon_{e \tau}$ and hence spectrum information can be useful
for removal of these degeneracies. 

\begin{figure}[H] 
\begin{center}
        \begin{tabular}{lr}
   \includegraphics[height=5cm,width=5.7cm]{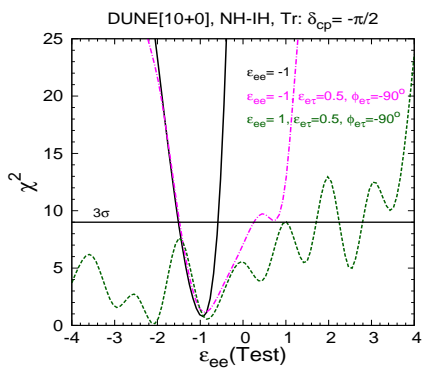}
        \end{tabular}
\end{center} 
\vspace{-4ex}
\caption{Hierarchy  $ \chi^{2} $  vs $ \epsilon_{e e} $ (Test)}
 \label{fig:hier_chisq}
\end{figure}

In fig.~(\ref{fig:hier_chisq}), we plot hierarchy $ \chi^{2} $ Vs $ \epsilon_{e e} $ (Test) to understand how the diagonal and off-diagonal NSI affect
the mass hierarchy sensitivity of DUNE while assuming NSI in nature. 
We have used General Long baseline Experiment Simulator (GLoBES) 
\cite{globes1} with additional tools from \cite{globes-nsi1}  in our numerical calculations. The experimental specifications and other numerical details are taken from \cite{Nath:2015kjg} except that this analysis is done for 40 kt detector mass.  The true values that we have considered are, $\sin^{2} \theta_{12} = 0.297$, $\sin^{2}2 \theta_{13} = 0.085$, 
$ \theta_{23} = 45^\circ $, $\dcp= - \pi/2$, $ \Delta m^{2}_{21} = 7.37 \times 10^{-5} eV^{2}$ and $ \Delta m^{2}_{31} = 2.50 \times 10^{-3} eV^{2}$  
\cite{Capozzi:2016rtj}.
Different true values of NSI parameters  are as mentioned in the 
fig.~(\ref{fig:hier_chisq}). 
We marginalize over $\theta_{13}$, $\dcp$, $\epsilon_{e \tau} $ and  $\phi_{e \tau}$. The remaining NSI parameters are taken to be zero.
The bounds on NSI parameters \cite{Biggio:2009nt,Ohlsson:2012kf} that occur 
in $P_{\mu e}$, eq.~(\ref{eq:prob}), at 90$\%$ C.L. are considered to be $|\epsilon_{e e}| \leq 4$, and  $\epsilon_{e \tau} \leq 3.0$ and the corresponding off-diagonal phase $\phi_{e \tau} = [-\pi,\pi]$. 

The black solid curve shows that there is no hierarchy sensitivity for 
$\epsilon_{e e} = -1$ in the absence of other NSI parameters because of 
the intrinsic degeneracy in the appearance channel. 
Here, the small non-zero $ \chi^{2} $ at $ \epsilon_{e e}  = -1$  
arises from the disappearance channel $P_{\mu \mu}$.  Because of the intrinsic nature of this degeneracy,
if hierarchy sensitivity is observed in DUNE and NSI exists in nature then 
certain ranges of NSI parameters will be ruled out. 
For instance assuming existence of only diagonal NSI, 
the range  $-1.6 < \epsilon_{e e} < -0.8$  will be ruled out if 
DUNE observes $3\sigma$ hierarchy sensitivity. For other true values 
of $\epsilon_{ee}$ the  corresponding degenerate parameter space will be 
excluded.
The magenta curve shows the hierarchy sensitivity for a non-zero true value 
of $\epsilon_{e \tau} = 0.5$ and $\phi_{e \tau} = -\pi/2$.
We find that for this the global minima comes at $\epsilon_{ee} = -0.8$
and there is no hierarchy sensitivity. \\
\par Note that there is also a local minima for $\epsilon_{ee} = 1$, 
where we observed an intrinsic degeneracy in fig. (\ref{fig:prob_ee_45}).
The reasons for which we do not get the global minima 
at this value are (i) marginalization over the phase $\phi_{e \tau}$, which 
had been kept fixed in fig. (\ref{fig:prob_ee_45}), (ii) the broadband 
nature of the beam at DUNE.
The green dotted curve depicts the hierarchy sensitivity for a different 
true value of $\epsilon_{ee} = 1$. The global 
minima comes at $\epsilon_{ee} = -2$ as well as a very close minima
near $\epsilon_{ee} = -1$ and several other local minima.
Here the global minima exhibits the most general form of 
the degeneracy: 
$P^{NH}_{\mu e}(\epsilon_{ee}, \dcp, \epsilon_{e \tau}, \phi_{e \tau}) = P^{IH}_{\mu e}(\epsilon_{ee}^\prime, \dcp^\prime, \epsilon_{e \tau}^\prime,\phi_{e \tau}^\prime)$.
However in this case 
$2\sigma$ sensitivity can be achieved by DUNE if the region 
$\epsilon_{ee} < 0.8$ is excluded.    

Our study shows that in presence of true non-zero NSI 
parameters DUNE will not have any hierarchy sensitivity if marginalized over 
the model independent ranges of NSI parameters.
If however, we use the more restrictive model dependent bounds then
the hierarchy sensitivity of DUNE can improve considerably. 
For instance assuming only diagonal NSI and the model dependent bound
$ -0.9 < \epsilon_{ee} < 0.75$ \cite{Biggio:2009nt}, $3\sigma$ hierarchy sensitivity can be achieved for $ -0.7 < \epsilon_{ee} $ and it can be very large for 
higher values of $\epsilon_{ee}$ \cite{Masud:2016gcl}.\\

\textbf{Implications of $\epsilon_{ee}$ $(=\epsilon_{ee}^\oplus) =-1$ in the Sun :}
 Neutrinos, while travelling through the matter of varying density possibly undergo resonant flavor transitions through Mikheyev–Smirnov–Wolfenstein (MSW) effect. This resonance phenomenon accounts for the flavor transitions of solar neutrinos while they propagate from the core to the surface of the Sun. \\
\par The solar neutrino evolution equation can be written in an effective two flavor model under one-mass scale dominance (OMSD) approximation $|\Delta m^2_{31}| \rightarrow \infinity$ as :
\begin{equation}
  i\frac{d}{dr}
  \begin{pmatrix}
    \nu_e\\
    \nu_\mu
  \end{pmatrix}
  = H_{eff}
  \begin{pmatrix}
    \nu_e\\
    \nu_\mu
  \end{pmatrix}
  \label{eq:sol-evol}
\end{equation}
where $r$ is the coordinate along the neutrino trajectory and $H_{eff}$ is the effective hamiltonian given by sum of vacuum, standard matter and NSI parts as 
\begin{equation} 
H_{eff}=U_{12}
\left[ 
\begin{array}{cc}
       0   & 0            \\
       0   &\Delta_{21}  
\end{array} 
\right]U_{12}^{\dagger} 
 + \sum_f V_f
  \begin{pmatrix}
    c_{13}^2 \delta_{ef}-\Eps_D^{f\hphantom{*}} & \Eps_N^f \\
    \hphantom{+} \Eps_N^{f*} & \Eps_D^f
  \end{pmatrix}
\label{eq:H-eff}
\end{equation}
where $U_{12}=\left[
\begin{array}{cc}
c_{12}  &  s_{12} \cr 
-s_{12} &  c_{12} \cr 
\end{array}
\right]$ , $\Delta_{21} \equiv \Delta
m^2_{21}/(4E)$ and $V_f = V_f(r) \equiv \sqrt 2 G_F N_f(r)$.
Here, the usual MSW effect induced by standard matter potential $V_e(r)$ is in the $(1,1)$ element of the second term in $H_{eff}$.
The diagonal $\epsilon_D^f$ (real) and off-diagonal $\epsilon_N^f$ (complex)
NSI parameters are related to $\epsilon_{\alpha\beta}$ 
by ~\cite{Kuo:1986sk,Gonzalez-Garcia:2013usa},
%
%
\begin{eqnarray}
\label{eq:eps_D}
\hspace{-1.4cm}\epsilon_D^f & = &
 c_{13} s_{13} \text{Re} \left[ \text{e}^{i\delta_{CP}} 
 ( s_{23} \, \epsilon_{e\mu}^f
       + c_{23} \, \epsilon_{e\tau}^f ) \right]
 \nonumber    \\
 &    - &( 1 + s_{13}^2 ) c_{23} s_{23} 
 \text{Re} ( \epsilon_{\mu\tau}^f)
\nonumber    \\
   &  - & \frac{c_{13}^2}{2} \big( \epsilon_{ee}^f - \epsilon_{\mu\mu}^f \big)
    + \frac{s_{23}^2 - s_{13}^2 c_{23}^2}{2}
    \big( \epsilon_{\tau\tau}^f - \epsilon_{\mu\mu}^f \big) \,, \nonumber \\
\hspace{-1.4cm} \epsilon_N^f 
&=&
 c_{13} \big( c_{23} \, \epsilon_{e\mu}^f - s_{23} \, 
\epsilon_{e\tau}^f \big)
 + s_{13} e^{-i\delta_{CP}} \times  \nonumber    \\
 & &\left[   s_{23}^2 \, \epsilon_{\mu\tau}^f - 
c_{23}^2 \, \epsilon_{\mu\tau}^{f \ast}
   + c_{23} s_{23} \big( \epsilon_{\tau\tau}^f - \epsilon_{\mu\mu}^f \big)
  \right]. \hskip 1.0cm 
\end{eqnarray}
%
The survival probability of solar neutrinos under OMSD approximation is given by :
\begin{equation}
P_{ee}=s_{13}^4+c_{13}^4 P_{eff}.
\end{equation}
where $P_{eff}$ can be obtained by solving eq.~(\ref{eq:sol-evol}) in the effective 2-flavor system.
By diagonalizing the effective hamiltonian in eq.~(\ref{eq:H-eff}) one can obtain the effective mixing angle $\theta_M$ in the matter as :
{\small \begin{equation} \label{eq:tan}
\tan2\theta_M = \frac{2(\Delta_{21}\sin2\theta_{12}+\sum_f V_f  \epsilon_N^f)}{2\Delta_{21}\cos2\theta_{12}-\sum_f V_f (c_{13}^2\delta_{ef}- 2\epsilon_D^f)}
\end{equation} }
It can be noted that the eq.~(\ref{eq:tan}) reduces to the effective mixing angle $\tan2\theta$ of a neutrino encountering only the standard matter potential when the NSIs cease to exist i.e. $\epsilon_{\alpha\beta}=0$. 


\par In the previous section, while considering the presence of only diagonal NSI parameter $\epsilon_{ee}$ we have taken a special case of $\epsilon_{ee} = -1$ for the Earth. Below, we have explored how this choice of NSI parameters would affect the MSW resonance in the Sun. \\
For the matter composition of the Earth the average of $Y_n$ is given by the PREM model as $Y_n=1.051$.
Thus from eq.~(\ref{eq:eps-ab}) we have 
\begin{equation}
\epsilon_{ee}=\epsilon_{ee}^e+3 \epsilon_{ee}^u+3\epsilon_{ee}^d~.
\label{eq:eps-earth}
\end{equation}
Whereas, in the Sun, $Y_n(r)$ varies from $1/2$ in the core to $1/6$ at the surface. Relevant expression for $\epsilon_{ee}^\odot$ can be obtained by substituting these values in eq.~(\ref{eq:eps-ab}).

For instance, let us consider $\epsilon_{ee}^u=-1/3$ and $\epsilon_{ee}^d=0$ which are within 1$\sigma$ 
allowed region of CHARM  results \cite{CHARM}.
Substituting $\epsilon_{ee}^u=-1/3$ and $Y_n = 1/2$ (at the core), in eq.~(\ref{eq:eps-ab}) gives 
\begin{equation}
\epsilon_{ee}^\odot=(2+ 1/2) \epsilon_{ee}^u
 \equiv - \frac{5}{6}~.
\label{eq:eps-sun1}
\end{equation}
Thus, eq.~(\ref{eq:eps_D}) reduces to 
\begin{equation}
\epsilon_D^u= \frac{-c_{13}^2}{2} \epsilon_{ee}^u = \frac{c_{13}^2}{6} \,
\hspace{0.2cm};
\hspace{0.2cm}
\epsilon_N^u= 0.
\end{equation} 
%
Now, by substituting the above equations in eq.~(\ref{eq:tan}) one can obtain a simplified form of
{\small \begin{eqnarray} \label{eq:tan1}
\tan2\theta_M &=& \frac{2\Delta_{21}\sin2\theta_{12}}{2\Delta_{21}\cos2\theta_{12}-V_e c_{13}^2 / 6},\\
&=& \frac{\tan2\theta}{1-\frac{N_e}{N_{e,res}}}~.
\end{eqnarray} }
Thus, MSW resonance in Sun occurs when the condition 
\begin{equation}
\Delta_{21}\cos2\theta_{12} = V_e c_{13}^2 / 12
\label{eq:res-sun} \,,
\end{equation}
is satisfied.
The electron density at resonance is,
\begin{equation}
N_{e,res}=\frac{3\Delta m^2_{21}\cos2\theta_{12}}{\sqrt{2}G_F E c_{13}^2 }~.
\end{equation}
We have observed that for neutrino energy $E \geq 7.2 $ MeV MSW resonance occurs in the Sun, i.e. the matter mixing angle is $ \cos 2\theta_{M} \sim -1 $ and the corresponding survival probability is $ P_{ee} \sim  0.3 $ 
for $ \Delta m^2_{21} \sim 4 \times 10^{-6}eV^2$ and Large Mixing Angle solution with $\sin^2 \theta_{12} = 0.3$.
This analysis can be extrapolated to the case of non-zero off-diagonal NSI parameters (say $|\epsilon_{e\tau}^\odot| \neq 0$). However, this would require a thorough study of non-standard neutrino interactions in the Sun which is beyond the scope of this paper. 

\textbf{Conclusions :} 
In this paper, we make the striking observation that if the 
parameter $\epsilon_{ee}=-1$, the standard matter effect gets cancelled  
by the NSI effects and the probability $P_{\mu e}$ is just the vacuum 
oscillation probability in absence of off-diagonal NSI parameters. 
 If in addition, the CP phase $\delta_{CP} = \pm \pi/2$ 
one gets an intrinsic hierarchy degeneracy which cannot be removed
even if both $\epsilon_{ee}$ and $\dcp$ are known precisely. 
This result acquires more relevance in the light of the preference 
of T2K  data and global oscillation analysis 
results implying  $\dcp \sim -\pi/2$. 
Although this is a special case of the 
generalized hierarchy degeneracy  
$\epsilon_{ee} \rightarrow -\epsilon_{ee} - 2 $,
$\dcp \rightarrow \pi-\dcp$, considered in \cite{Coloma:2016gei}, 
for any other value of $\epsilon_{ee}$ and $\dcp$,  a 
precise measurement of these parameters will  alleviate the
degeneracies. However for $\epsilon_{ee} = -1$ together with $\dcp = \pm \pi/2$ that is not true.   
To the best of our knowledge 
this particular point  has not been highlighted 
before. This conclusion, being independent of baseline 
and energy,  can seriously impact the 
hierarchy sensitivity of the DUNE experiment.  

The matter effects can reappear, 
if for instance, the off-diagonal NSI parameter $\epsilon_{e\tau}$ is included. 
But we show that for this case 
the intrinsic hierarchy degeneracy is  transported  from 
$\epsilon_{ee} = -1$ to a different 
value depending on the off-diagonal
NSI parameters as well as energy. 
Moreover the  uncertainty in the magnitude 
and phase of the off-diagonal NSI parameters can give rise to 
additional degeneracies affecting the hierarchy sensitivity.
A more precise knowledge of the parameter $\epsilon_{ee}$, can however,  
give rise to an enhanced sensitivity provided $\epsilon_{ee} \neq -1$.       
%
%
This underscores the importance of 
independent measurement of the NSI parameters from non-oscillation 
experiments  
\cite{Akimov:2015nza}. Furthermore, we also discussed the implications of $\epsilon_{ee} = -1$ 
(in the earth) on the matter effect in the Sun.


\textbf{Acknowledgement:} Authors thank Monojit Ghosh for his help with the code and many useful discussions. 


\end{document}